\documentclass[aps,prab,groupedaddress,floats,reprint]{revtex4-2}

\usepackage[english] {babel}
\usepackage{amsfonts,graphicx,soul,natbib,mathrsfs,dsfont}

\usepackage{amssymb,amsmath,comment}
\usepackage{cancel}
\usepackage{color}

\usepackage[colorlinks=true]{hyperref}

\usepackage{graphicx}% Include figure files
\newcommand{\g}{\gamma}
\newcommand{\G}{\Gamma}
\renewcommand{\d}{\delta}
\newcommand{\e}{\varepsilon}

\newcommand{\p}{\pi}
\renewcommand{\r}{\rho}

\newcommand{\ph}{\varphi}
\renewcommand{\c}{\chi}

\renewcommand{\o}{\omega}

\newcommand{\vc}[1]{\boldsymbol{\mathrm{#1}}}

\newcommand{\skob}[1]{\left( #1\right)}
\newcommand{\pskob}[1]{\left| #1\right|}
\newcommand{\sqskob}[1]{\left[ #1\right]}

\newcommand{\parl}{\parallel}      %Gparallel symbol

\newcommand{\til}[1]{\Tilde{#1}} %Tilde 
\begin{document}

\author{L. A. Shaposhnikov}%
\email{leon.shaposhnikov@metalab.ifmo.ru}%
\affiliation{School of Physics and Engineering,
ITMO University, St. Petersburg, Russia 197101}%
\author{S. S. Baturin}%
\email{s.s.baturin@gmail.com}%
\affiliation{School of Physics and Engineering,
ITMO University, St. Petersburg, Russia 197101}%

\date{\today}

\begin{abstract}
We present the wakefield conformal mapping technique that can be readily applied to the analysis of the radiation generated by an ultra-relativistic particle in the step transition and a collimator. We derive simple analytical expressions for the lower and upper bounds of both longitudinal and transverse wake potentials. We test the derived expressions against well-known formulas in several representative examples. The proposed method can greatly simplify the optimization of collimating sections, as well as become a useful tool in the shape optimization problems. 
\end{abstract}

\title{Bounds on geometric wakefields in collimators and step transitions of arbitrary cross sections.}

\maketitle

%==============================================================
%==============================================================

\section{Introduction}

%==============================================================
%==============================================================

When a relativistic charged particle passes near an inhomogeneity, it emits electromagnetic fields that inevitably interact with particles that follow the source of the radiation. Such interactions by radiation of different types are generalized into a convenient framework known as wakefield interactions \cite{Chao,Zotter}. The analysis and computation of wakefields of various types - geometric, resistive wall \cite{Bane}, dielectric \cite{NG,rosen,rtg}, and even metamaterial \cite{wp} - are the cornerstone of the design of any modern accelerator, be it a conventional machine like CLIC \cite{CLIC}, ILC \cite{ILC}, LCLS-II \cite{LCLS} etc. or a wakefield accelerator like the A-STAR \cite{Zhol}. 

The wakefield generated in the transition of various components of the vacuum system (geometric wakes) and the related beam dynamics effects were among the first to be studied. Generally, these studies concentrated on particular geometries of the vacuum chambers and were initially restricted to circular cross sections. As accelerator technology advanced, researchers had to take into account more complex cross-sectional shapes, particularly in the design of the transition sections from the beam transport tube to the undulator in the case of X-FELs.

Geometric wakes may be essential for the design of modern collinear wakefield accelerators (CWAs). For example, in the A-STAR project \cite{Zhol,Zhol2}, when transitioning from a small aperture of the corrugated vacuum chamber \cite{SiyCor} to the transition section \cite{Siy}, witness and driver beams may be significantly affected by wake influence, which could be reduced through proper optimization. Direct numerical treatment \cite{ECHO,Vay_2012} of this issue is the most accurate approach, yet optimization procedures for bunches with lengths of around 1 $\mu$m can be computationally expensive. Calculating the short-range wake requires a spatial mesh size that is a fraction of the bunch length, making submillimeter and micrometer-scale bunches a challenge.

A comprehensive study of geometric wakes within the optical approximation is considered in\cite{Opt1,Opt2}, where several exact analytical results on the wake potential (both longitudinal and transverse) have been obtained for step transitions, collimators and slits. While the approach of Ref.\cite{Opt1} provides a very powerful tool for the evaluation of geometric wakes, further improvement and simplification of this method are of additional value, especially in view of new possible applications to the CWA. In the present study, we follow the conformal mapping approach introduced in Refs.\cite{myPRL,mySTAB} for the steady-state wake in structures of arbitrary cross sections. This method has proven useful in several advanced accelerator applications, such as flat beams Refs.\cite{planar,SprPRL} and the recently introduced L-shape corrugated structure \cite{Lshape}. We start from the seminal idea of Heifets and Kheifets \cite{HfKf} for the circular collimator and combine it with the conformal mapping approach \cite{mySTAB} and concepts of optical approximation of Ref.\cite{Opt1}. Based on the energy balance equation, we express the upper and lower bounds of the wake potential by the difference of the electromagnetic self-energy of the two-particle bunch in the incoming and outgoing beam pipes and independently arrive at the same cross-integral as in Ref.\cite{Opt1}, which was derived using the indirect integration technique \cite{ii1,ii2}. We go one step further and evaluate this integral explicitly for the case of arbitrary cross sections of the collimator tubes. We show that upper and lower bounds for the exact Green's function for the longitudinal wake potential are expressed by a logarithm of the ratio of the corresponding conformal maps. The latter is a natural generalization of the well-known result for a round pipe. By means of the Panofsky-Wenzel theorem \cite{PW} we relate corresponding bounds on the Green's function for the transverse wake potential to the corresponding conformal maps and their derivatives. We conclude with a series of simple examples and transitions that reproduce known results.   

%==============================================================
%==============================================================

\section{Basic equations}

%==============================================================
%==============================================================

We consider the Maxwell system in CGS units in vacuum
\begin{align}
\nabla\times\mathbf{E}&=-\frac{1}{c}\frac{\partial \mathbf{H}}{\partial t }, \nonumber \\ 
\nabla\times\mathbf{H}&=\frac{4\pi}{c}\mathbf{j}+\frac{1}{c}\frac{\partial \mathbf{E}}{\partial t }, \\ \nonumber
\nabla\cdot\mathbf{E}&=4\pi\rho, \\
\nabla\cdot\mathbf{H}&=0. \nonumber
\end{align}
We assume that the particle is moving along the z-axis.  In the ultra-relativistic limit the current produced by the particle and its charge density are connected through:
\begin{align} 
\label{src}
|\mathbf{j}|=j_z= c \rho(x,y,z-ct).
\end{align}
First, we introduce a new coordinate $\zeta=ct-z$ and split the Maxwell system into two parts:
\begin{align}
\label{Max1}
[\nabla\times\mathbf{E}]_\bot&=-\frac{\partial \mathbf{H_\bot}}{\partial \zeta }, \nonumber \\ 
[\nabla\times\mathbf{H}]_\bot&=\frac{\partial \mathbf{E_\bot}}{\partial \zeta };
\end{align}
and
\begin{align}
\label{Max2}
\left[\nabla\times\mathbf{E} \right]_z&=-\frac{\partial H_z}{\partial \zeta }, \nonumber \\ 
\left[\nabla\times\mathbf{H} \right]_z&=4\pi\rho+\frac{\partial E_z}{\partial \zeta }, \\ \nonumber
\nabla_\bot\cdot\mathbf{E}_\bot&=4\pi\rho+\frac{\partial E_z}{\partial \zeta }, \\
\nabla_\bot\cdot\mathbf{H}_\bot&=\frac{\partial H_z}{\partial \zeta }. \nonumber
\end{align}
Here the $\bot$ -symbol indicates field components orthogonal to the z-axis and $\nabla_\bot$ - is the nabla operator in the plane orthogonal to the z-axis.

Next, as in Ref.\cite{mySTAB}, we combine Eqs.\eqref{Max2} as 
\begin{align}
\label{Max3}
\nabla_\bot\cdot\mathbf{E}_\bot+i\left[\nabla\times\mathbf{E}\right]_z&=4\pi\rho+\frac{\partial E_z}{\partial \zeta }-i\frac{\partial H_z}{\partial \zeta }, \\
\nabla_\bot\cdot\mathbf{H}_\bot+i \left[\nabla\times\mathbf{H}\right]_z&=i\left(4\pi\rho+\frac{\partial E_z}{\partial \zeta }-i\frac{\partial H_z}{\partial \zeta }\right). \nonumber
\end{align}
Finally, we introduce complex functions
\begin{align}
\label{c_fl}
e&=E_x+iE_y, \nonumber \\
h&=H_x+iH_y, 
\end{align}
and note that ultra-relativistic particle inside a perfectly conducting pipe do not radiate, consequently $E_z=H_z=0$ and the last terms in \eqref{Max3} vanish. Therefor, in complex notation of Eq.\eqref{c_fl} one may reduce Eqs.\eqref{Max3} to 
\begin{align}
\label{c_maxl}
\frac{\partial e}{\partial \omega}&=2\pi \rho. 
\end{align}
Here $\omega=x+iy$. Since $h=ie$ (see Appendix \ref{app:2} for details), we further consider only the equation for $e$.
Eq.\eqref{c_maxl} must be supplemented with the boundary condition at a perfect conductor (PEC) that reads
\begin{align}
    \Re[e \tau^*]=0.
\end{align}
Here, $\tau=\tau_x+i\tau_y$ is the tangential vector to the pipe surface in the cross section orthogonal to the $z$-axis, the asterisk represents the complex conjugation.

We introduce potential $\ph$ as 
%\commentLeon{Here also only one *} 
\begin{align}
\label{eq:potd}
    e=-2\frac{\partial \ph^*}{\partial \omega^*}.
\end{align}

It is apparent that $\ph$ is a solution to the 2D Poisson equation with the Dirichlet boundary condition on $\Gamma_0$ - the curve that encloses the vacuum channel of the pipe. We note that  $\Delta_\perp=4\partial_\o\partial_\o^*$ and write 
%\commentStas{In this definition '2' is missing, I think...}
\begin{align}
\label{eq:Php}
    &\Delta_\perp \ph = -4\pi \rho, \\  \nonumber
    &\ph|_{\Gamma_0}=0.
\end{align}
\begin{figure}[t]
    \label{fig: conformal map}
    \includegraphics[width=0.5\textwidth]{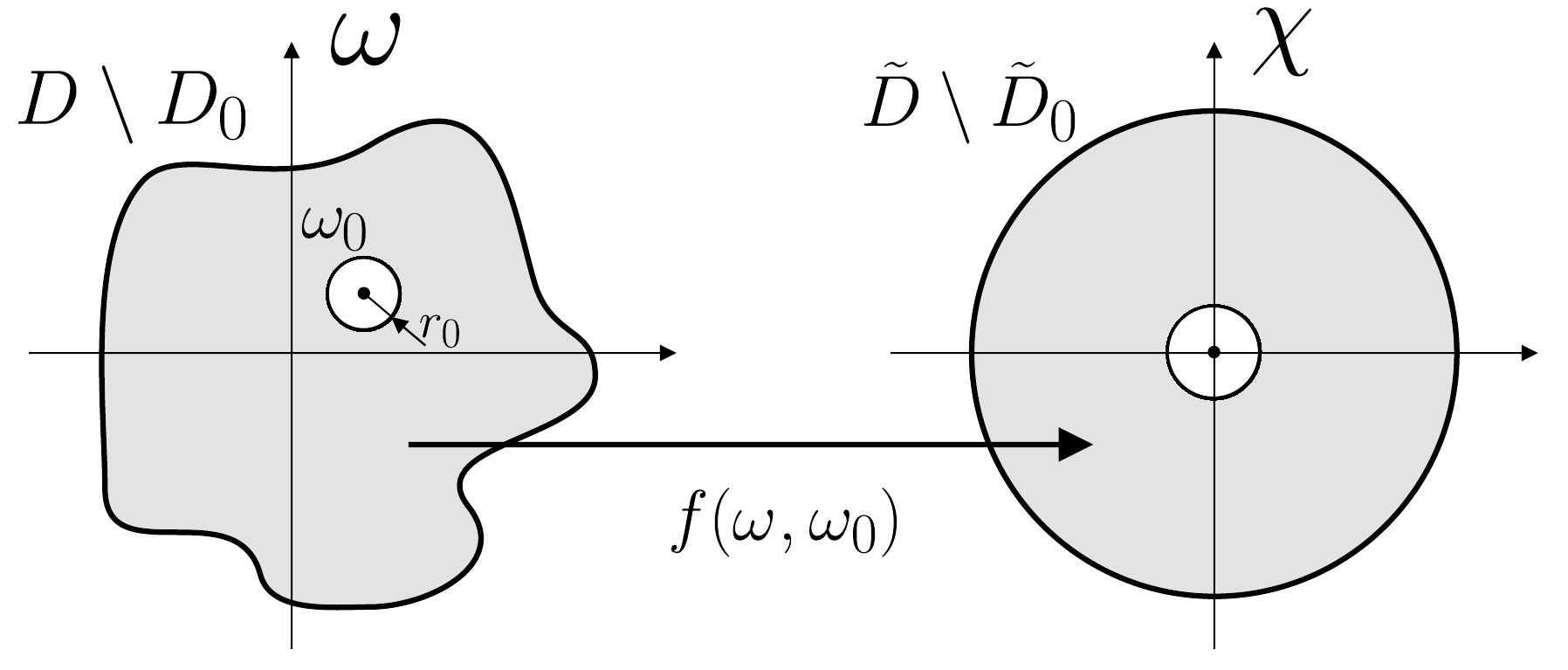}
    \caption{Action of the conformal map $f(\o,\o_0)$ that transforms area $D$ into unit disk, while point $\o_0$ is mapped to the center of this disk.}
\end{figure}

Solution to the problem \eqref{eq:Php} above is well known (see for instance Refs.\cite{shabat,silverman}) and reads

\begin{align}\label{eq: potenrial in terms of green function lnf}
    \ph=-2\int dx_0 dy_0~ \rho(x_0,y_0,\zeta) \ln \left|f(\omega,\omega_0) \right|.
\end{align}

Here, $f(\omega,\omega_0)$ is the conformal mapping function that gives the mapping of the $D$ -region onto a disk such that the point $\omega_0$ is mapped to the center of a disk. 
The electric field is recovered with the help of Eq.\eqref{eq:potd} and reads

\begin{align}
\label{eq:esf}
    e=2\int dx_0dy_0~ \rho(x_0,y_0,\zeta) \left[\frac{f'(\omega,\omega_0)}{f(\omega,\omega_0)}\right]^*,
\end{align}

where the prime symbol denotes the derivative by $\omega$.

%==============================================================
%==============================================================

\section{Energy evaluation}

%==============================================================
%==============================================================

We start from the energy balance equation

\begin{align}
\label{eq:PoThM}
    -\frac{\partial}{\partial t} \int_V w dV = \int_{\partial V} \vc{S}\cdot\mathrm{d}\vc{A}+\int_{V} \vc{j}\cdot \vc{E} dV. 
\end{align}
We assume that the total energy $W=\int_V w dV$ is conserved. This implies that 
\begin{align}
\label{eq:enrg}
    W(t=-\infty)=W(t=\infty).
\end{align}
The beam and the radiation travels along the $z$ axis, consequently only $z$-component of the Poynting's vector does not vanish.
After integrating Eq.\eqref{eq:PoThM} over time and accounting for the condition Eq.\eqref{eq:enrg} we get
\begin{align}
\label{eq:PoTh}
   &\int\limits_{-\infty}^{\infty}\int\limits_{D_A} S_zdxdydt+\int\limits_{-\infty}^{\infty}\int\limits_{D_B} S_zdxdydt\\ \nonumber &+\int\limits_{-\infty}^{\infty}\int\limits_{V} j_z E_z dVdt =0.
\end{align}
The last term is the energy lost/gained by the bunch due to the interaction $U_{\mathrm{rad}}$ and the first two terms are the total fluxes $\Phi_A$ and $\Phi_B$ though the cross sections $D_A$ and $D_B$ correspondingly.

%We assume that both the cross section $D_A$ and $D_B$ are far enough from the interaction point such that all transition processes that arise from the collimating section are settled. 

The flux though the cross section $A$ consists of the incoming flux of the source and outgoing flux of the scattered field. The flux through the cross section $B$ includes the outgoing flux of the source and the remaining part of the scattered field flux.   
With the help of Eq.\eqref{c_fl} one may write for the Poynting vectors of the source
\begin{align}
\label{eq:Sz}
    S_z=\frac{c}{4\pi}\left[\vc{E}\times\vc{H}\right]_z=\frac{c}{4\pi}\Im[e^*h]
\end{align}
Recalling the connection $h=ie$ we get
\begin{align}
    S_z=\frac{c}{4\pi}|e|^2,
\end{align}
and thus
\begin{align}
\label{eq:flux}
    \Phi_{A,B}^{\mathrm{sl}}=\frac{c}{4\pi} \int\limits_{-\infty}^{\infty}\int\limits_{D_{A,B}} |e|^2 dx dy dt.
\end{align}
It is convenient to rewrite Eq.\eqref{eq:PoTh} as
\begin{align}
\label{eq:radgen}
 U_{\mathrm{rad}}=-\Phi_{B}^{\mathrm{sl}}+\Phi_{A}^{\mathrm{sl}}-\Phi^{\mathrm{scat}},   
\end{align}
where we have accounted for the sign of the fluxes and introduced $\Phi^{\mathrm{scat}}$ - the total flux of the scattered field (though both the cross section $D_A$ and $D_B$. The normal vector is pointed inside the volume enclosed by the beam pipes and cross sections $A$ and $B$. Negative sign in-front of the scattered field flux accounts for the fact that the scattered field propagates outside the volume and fluxes through both cross sections $A$ and $B$ are negative.
\begin{figure}[t]
\includegraphics[width=0.5\textwidth]{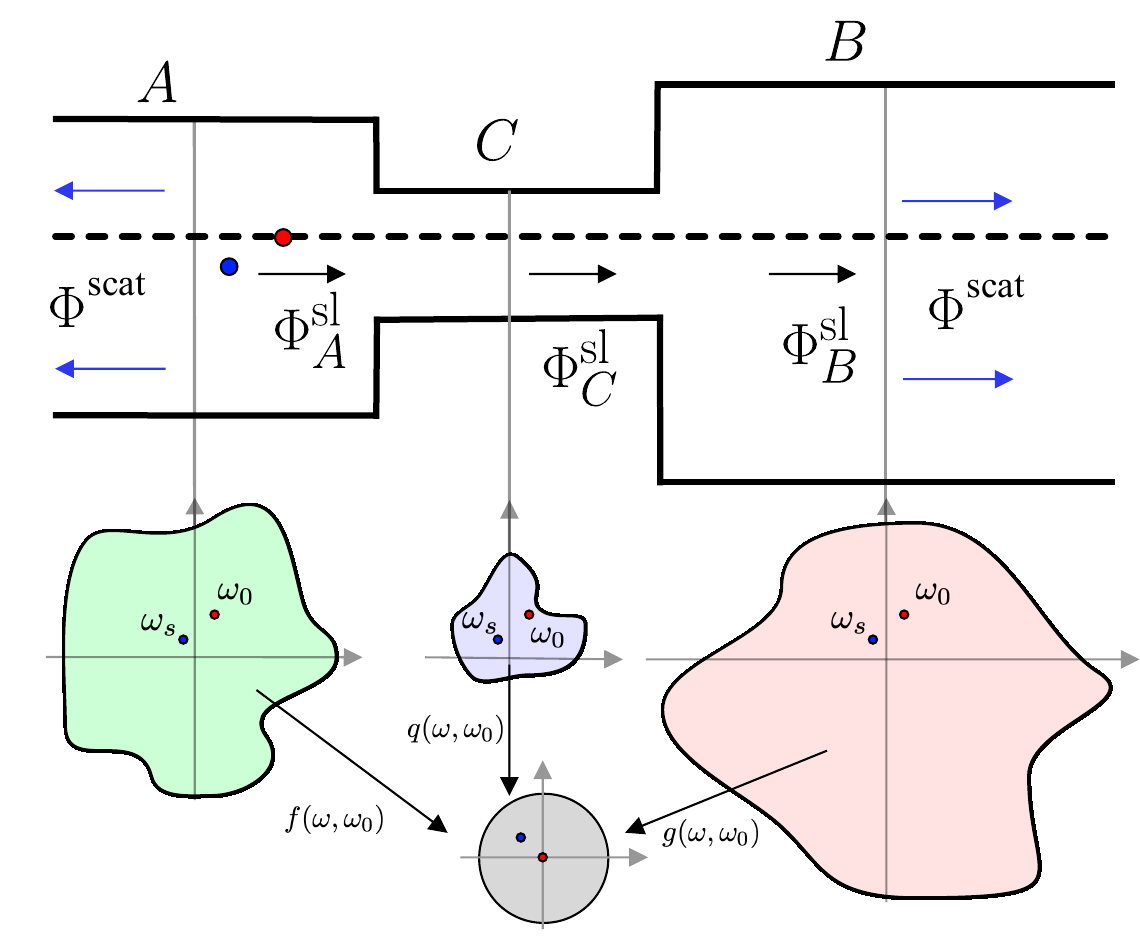}
    \caption{Sketch of the arbitrary step collimator. \label{Fig:2}}
\end{figure}
Eq.\eqref{eq:radgen} can be used right away to estimate the energy gain in the step-in transition. Replacing cross section $B$ with cross section $C$ (see Fig.\ref{Fig:2}) in Eq.\eqref{eq:radgen} we get the upper bound for the energy gain $U_{\mathrm{rad}}^{s.in}$ in the pure step-in transition.
\begin{align}
\label{eq:radsin}
    U_{\mathrm{rad}}^{\mathrm{s.in}}<\Phi_{A}^{\mathrm{sl}}-\Phi_{C}^{\mathrm{sl}},
\end{align}
a known phenomenon (see for instance Ref.\cite{HfKf}). It should be noted that the flux $\Phi^{\mathrm{scat}}$ satisfies the same inequality as follows from Eq.\eqref{eq:radgen} under the assumption $U_{rad}>0$, namely.
\begin{align}
\label{eq:sin}
    \Phi^{\mathrm{scat}}_{\mathrm{s.in}}<\Phi_{A}^{\mathrm{sl}}-\Phi_{C}^{\mathrm{sl}}.
\end{align}
One of the important observations made in Ref.\cite{HfKf} is that $\Phi^{\mathrm{scat}}$ is the same for both step-in and step-out transition of the same configuration. This is a direct consequence of the parity-time symmetry of the problem and the application of the Lorentz reciprocity theorem. Thus, following Ref.\cite{HfKf} we may write for the step-out
\begin{align}
\label{eq:sout}
    \Phi^{\mathrm{scat}}_{\mathrm{s.out}}<\Phi_{B}^{\mathrm{sl}}-\Phi_{C}^{\mathrm{sl}}.
\end{align}
To get the estimate for the upper bound of the energy loss we account for Eq.\eqref{eq:sin}, Eq.\eqref{eq:sout} and note that the total flux $\Phi^{\mathrm{scat}}$ for the collimator is the sum of step-in and step-out fluxes. Consequently, we have
\begin{align}
\label{eq:fltot}
    \Phi^{\mathrm{scat}}=\Phi^{\mathrm{scat}}_{\mathrm{s.in}}+\Phi^{\mathrm{scat}}_{\mathrm{s.out}}<\Phi_A^{\mathrm{sl}}+\Phi_B^{\mathrm{sl}}-2\Phi_C^{\mathrm{sl}}.
\end{align}
Under the assumption of $U_{\mathrm{rad}}<0$ in a collimator and combining Eq.\eqref{eq:fltot} with Eq.\eqref{eq:radgen} we get
\begin{align}\label{eq:apprd}
    |U_{\mathrm{rad}}|<2\Phi_{B}^{\mathrm{sl}}-2\Phi_{C}^{\mathrm{sl}}.
\end{align}

We note that provided upper bound rapidly approaches the exact value for the $|U_{\mathrm{rad}}|$ when conditions of the optical approximation are met:
\begin{align}
\label{eq:opt}
    \lambda\ll b,~l\ll \frac{b^2}{\lambda}.
\end{align}
Here $\lambda$ is the reduced wavelength of the radiation, $l$ is the length of an obstacle, and $b$ is the minimal
cross section size of the beam pipe.

The first condition indicates that the obstacle must be much larger than the reduced wavelength of the radiation. The second condition requires the object to be short compared to the catch-up distance. More information on the optical approximation can be found in Ref.\cite{Opt1}.
We note that for the short bunches this approximation provides good quantity agreement between analytical results and simulations \cite{Opt1,Opt2,HfKf}. 

Dropping $\Phi^{\mathrm{scat}}$ in Eq.\eqref{eq:radgen} on can get the lower bound for the $|U_{\mathrm{rad}}|$, that we combine with Eq.\eqref{eq:apprd} and finally get

 \begin{align}
 \label{eq:radfin2}
     \Phi_{B}^{\mathrm{sl}}-\Phi_{A}^{\mathrm{sl}}<|U_{\mathrm{rad}}|<2\Phi_{B}^{\mathrm{sl}}-2\Phi_{C}^{\mathrm{sl}}.
 \end{align}

We point out that both bounds are firm and valid regardless of the approximation and are derived from the estimates on $\Phi^{\mathrm{scat}}$, however, both bounds hold only if one knows for a fact that the beam experiences net energy loss when passing through a collimator.
 
Thus, the lower and upper bounds for the energy loss in a collimator transition can be estimated using the differences of the charge self-energies calculated in different cross sections. 

It is worth pointing out that if the scattered energy is dissipated or trapped inside the volume, then an additional energy term that effectively decreases the scattered field flux must be accounted for, this, however, does not affect the upper bound and the inequality given by Eq.\eqref{eq:radfin2} remains valid.  

%==============================================================
%==============================================================

\section{Loss factor and Wake potential}

%==============================================================
%==============================================================

%==============================================================
\subsection{Loss factor of a single short bunch \label{sec:sp}}
%==============================================================

We start from a single, short $\sigma_z\ll a$ (where $a$ - is the characteristic size of the smallest aperture), pencil-like Gaussian bunch 
\begin{equation}
\label{eq:sinp}
\r(x,y,\zeta)=Q\frac{\d\skob{x-x_0}\d\skob{y-y_0}}{\sqrt{2\pi}\sigma_z} \exp\left[{-\frac{\zeta^2}{2\sigma_z^2}}\right].
\end{equation}
Here, $Q$ is the total charge and $\sigma_z$ is the RMS bunch length. By plugging Eq.\eqref{eq:sinp} into the Eq.\eqref{eq:esf} we get

\begin{equation}
    e=2Q\frac{\exp\left[-\frac{\zeta^2}{2\sigma_z^2} \right]}{\sqrt{2\p}\sigma_z}\left[\frac{f'(\o,\o_0)}{f(\o,\o_0)} \right]^*,
\end{equation}
with $\o_0=x_0+iy_0$.
With the help of the Eq.\eqref{eq:sinp} from Eq.\eqref{eq:flux} we get
\begin{align}
\label{eq:flux_gauss}
    \Phi_{B,A}^{\mathrm{sl}}=\frac{4Q^2}{4\pi}&\int\limits_{-\infty}^{\infty}cdt \frac{\exp\left[-\frac{\zeta^2}{\sigma_z^2} \right]}{2\p\sigma_z^2}\times \\ \nonumber &\int\limits_{D_{A,B}} \frac{|f'(\o,\o_0)|^2}{|f(\o,\o_0)|^2} dx dy 
\end{align}
The first integral in Eq.\eqref{eq:flux_gauss} with $d\zeta=cdt$ evaluates to
\begin{align}
\label{eq:longI}
    \int\limits_{-\infty}^{\infty}d\zeta \frac{\exp\left[-\frac{\zeta^2}{\sigma_z^2} \right]}{2\p\sigma_z^2}=\frac{1}{2\sigma_z\sqrt{\pi}}.
\end{align}

The second integral over $D_{A,B}$ does not converge at the lower limit. To eliminate this divergence, we cut out the circle of the infinitesimally small radius $r_0$ with the center in the $\omega_0$, we call this area $D_0$. Next, we integrate over $D_{A,B} \setminus D_0$. 

We recall that switching from $\o$ to $\chi=f(\o,\o_0)=\tilde x+i \tilde y$ (see Fig.\ref{fig: conformal map}) results in the Jacobian in the transformation of the elementary square:
$d\tilde x d \tilde y=\pskob{f'(\o,\o_0)}^2dxdy$. 
Boundaries $\partial \tilde D_{A,B}$ and the boundary of the cut circle $\partial \tilde D_0$ in the $\chi$ plane are defined as
\begin{align}
&\partial \tilde D_{A,B}=\left\{ \vc{\tilde r} |\pskob{\vc{\tilde r}}=1  \right\}, \nonumber\\
&\partial \tilde D_0=\left\{ \vc{\tilde r} | \forall  \ph: f\skob{\o_0+r_0e^{i\ph},\omega_0}  \right\}.
\end{align}
We assume that $r_0$ is much smaller than any other characteristic size in our problem.  Thus, we expand $f\skob{\o_0+r_0e^{i\ph},\o_0}\approx f\skob{\o_0,\o_0}+f'\skob{\o_0,\o_0}r_0e^{i\ph}=f'\skob{\o_0,\o_0}r_0e^{i\ph}$. With this we can define $\tilde D_0$ as a disk with radius $r_0\pskob{f'\skob{\o_0,\o_0}}$ centered at the point $0$:
\begin{equation}
    \partial \tilde D_0=\left\{ \vc{\tilde r} |\pskob{\vc{\tilde r}}=r_0\pskob{f'\skob{\o_0,\o_0}}  \right\}.
\end{equation}

In the $\chi$ plane where $\tilde D_{A,B}$ is now a disk and the last integral in Eq.\eqref{eq:flux_gauss} can be easily evaluated. If $f$ is the mapping of the $D_A$ onto a unit disk we get

\begin{align}
\label{eq:intDA}
&\int\limits_{D_A\setminus D_0}\pskob{\frac{f'(\o,\o_0)}{f(\o,\o_0)}}^2dxdy=\int\limits_{\tilde D_A\setminus \tilde D_0}\frac{1}{|f(\o,\o_0)|^2}d\tilde x d \tilde y= \nonumber \\ & \int\limits_{r_0\pskob{f'(\o_0,\o_0)}}^{1}\frac{2\p}{\tilde r}d \tilde r=-2\p\ln \skob{r_0\pskob{f'(\o_0,\o_0)}}.
\end{align}
If now $g(\omega,\omega_0)$ is the mapping of the area $D_B$ the cross section of the outgoing pipe onto the unit disk such that $\omega_0$ - the transverse position of the pencil beam is mapped to the center of the disk we get 
\begin{align}
\label{eq:intDB}
&\int\limits_{D_B\setminus D_0}\pskob{\frac{g'(\varkappa,\varkappa_0)}{g(\omega,\omega_0)}}^2dxdy=\int\limits_{D'_B\setminus D_0'}\frac{1}{|g(\omega,\omega_0)|^2}d\tilde xd \tilde y= \nonumber \\ & \int\limits_{r_0\pskob{g'(\omega,\omega_0)}}^{1}\frac{2\p}{\tilde r}d\tilde r=-2\p\ln \skob{r_0\pskob{g'(\omega_0,\omega_0)}}.
\end{align}
From Eq.\eqref{eq:radfin2} combining Eq.\eqref{eq:flux_gauss}, Eq.\eqref{eq:longI}, Eq.\eqref{eq:intDA} and Eq.\eqref{eq:intDB} we evaluate the lower bound for the energy loss as: 
\begin{equation}\label{eq: single particle final energy}
    |U_{\mathrm{rad}}|> \frac{1}{2\sigma_z\sqrt{\pi}}\frac{Q^2}{2}4\ln \pskob{\frac{f'(\o_0,\o_0)}{g'(\o_0,\o_0)}}.
\end{equation}    
The loss factor $\kappa_\parallel=|U_{rad}|/Q^2$ is then
\begin{equation}
\label{eq:loss factor}
    \kappa_\parallel > \frac{1}{2\sigma_z\sqrt{\pi}}\frac{1}{2}4\ln \pskob{\frac{f'(\o_0,\o_0)}{g'(\o_0,\o_0)}}.
\end{equation}    
we observe that we recover familiar logarithmic term. Yet, the form of this term is now more general and accounts for the form factors of the corresponding pipes.

We note the factor $1/2$ in both Eq.\eqref{eq:loss factor} and Eq.\eqref{eq: single particle final energy}. 
Next, we consider a symmetric transition (when the incoming and outgoing pipes have the same cross section). We introduce a conformal mapping function $q(\o,\o_0)$, which maps the cross section C of the central part of the collimator onto a unit disk in such a way that the point of the particle $\o_0$ is mapped onto the center of the disk. With the help of Eq.\eqref{eq:radfin2} we arrive at a generalization of the formula for the loss factor of the arbitrary step collimator: 
%with equal incoming and outgoing pipes:
\begin{align}
\label{eq:loss factor col}
    \kappa_\parallel^{\mathrm{col}} < \frac{1}{2\sigma_z\sqrt{\pi}}4\ln \pskob{\frac{f'(\o_0,\o_0)}{q'(\o_0,\o_0)}}.
\end{align}     

Using the inequality Eq.\eqref{eq:radfin2}, the generalized formula for the beam with the longitudinal charge distribution $\rho_\parallel(s)$ passing through a three-pipe junction (a full step in and a full step out) %with the additional condition $D_A \subseteq D_B$ is as follows
\begin{align}
\label{eq:loss factor gen}
    \kappa_\parallel&>2||\rho_\parallel(s)||_{L^2}^2 \ln \pskob{\frac{f'(\o_0,\o_0)}{g'(\o_0,\o_0)}}, \nonumber \\
    \kappa_\parallel&< 4||\rho_\parallel(s)||_{L^2}^2\ln \left| \frac{q'(\o_0,\o_0)}{g'(\o_0,\o_0)} \right|.
\end{align}    
Where $||\rho_\parallel(s)||_{L^2}^2$ is the square of the $L^2$ norm of the longitudinal bunch distribution and is given by  
\begin{align}
||\rho_\parallel(s)||_{L^2}^2=\int\limits_{-\infty}^{\infty} |\rho_\parallel(s)|^2ds.    
\end{align}

%Note an additional approximation that is silently assumed. In the region where $r/\lambda>\g$ (here as before $\lambda$ is  the reduced wavelength of the radiation), the electromagnetic fields are exponentially small, and one must switch to the Fourier representation of the problem and truncate the inverse Fourier integral over $k\sim 1/\lambda$ to $\g/r$. If we assume that the bunch never couples to the modes with the $k \sim \gamma/a$, where $a$ is the characteristic size of the smallest pipe, then the inverse integral can be approximately replaced by $\infty$ and the analysis holds. Note also that for the limiting case $\gamma=\infty$ the analysis is exact. 

%==============================================================
%==============================================================

\subsection{Green's function and wake potential}

%==============================================================
%==============================================================

In this section we derive a two-point function and relate it to the wake potential.
We extend our analysis to the general case of two particles, one traveling behind the other at a distance $s$. 

The charge density in this case is
\begin{align}
\label{eq:dp}
\r=\r_1+\r_2=&Q_1\d\skob{x-x_0}\d\skob{y-y_0}\d\skob{\zeta}+\nonumber \\ &Q_2\d\skob{x-x_s}\d\skob{y-y_s}\d\skob{\zeta-s}.
\end{align}
With the help of Eq.\eqref{eq:dp} and Eq.\eqref{eq:esf} we get
\begin{align}
\label{eq:dpe}
e=e_1+e_2= 
&=Q_1\d(\zeta)\left[\frac{f'(\o,\o_0)}{f(\o,\o_0)} \right]^*+\nonumber \\
&+Q_2\d(\zeta-s)\left[\frac{f'(\o,\o_s)}{f(\o,\o_s)} \right]^*,
\end{align}
where $Q_1+Q_2=Q$ - the total charge of the two-particle bunch. 

We substitute Eq.\eqref{eq:dp} and Eq.\eqref{eq:dpe} into the Eq.\eqref{eq:PoTh} and get
\begin{align}
\label{eq: energy balance for 2 particles m}
&c Q_1\int\limits_{-\infty}^{\infty}E_z(x_0,y_0,z=ct,t)dt+ \\ \nonumber +&cQ_2\int\limits_{-\infty}^{\infty} E_z(x_s,y_s,z=ct-s,t)dt =-\Phi_{B}^{\mathrm{G}}+\Phi_{A}^{\mathrm{G}}-\Phi_{A}^{\mathrm{ref}}. 
\end{align}
With $\Phi_{A,B}^{G}$ given by
\begin{align}
    \Phi_{A,B}^{G}=&\frac{c}{4\pi}\int\limits_{-\infty}^{\infty}\int_{D_A,D_B}|e_1|^2dxdydt+\nonumber \\  &\frac{c}{4\pi}\int\limits_{-\infty}^{\infty}\int_{D_A,D_B}|e_2|^2dxdydt +\\&\frac{c}{4\pi}\int\limits_{-\infty}^{\infty}\int_{D_A,D_B}\skob{e_1e_2^*+e_2e_1^*}dxdydt. \nonumber
\end{align}
The first two terms correspond to the self interaction of each particle as was discussed in Sec.\ref{sec:sp}. The right-hand side of the energy balance equation \eqref{eq: energy balance for 2 particles m} consists of three terms the first term due to the causality (the longitudinal field created by the second particle at the position of the first particle is zero $E_{2,z}=0$) is just the loss of the first particle to the radiation
\begin{align}
    U_{\mathrm{rad}}^{(1)}=c Q_1\int\limits_{-\infty}^{\infty}E_{1,z}(x_0,y_0,z=ct,t)dt.
\end{align}
The second term in Eq.\eqref{eq: energy balance for 2 particles m} due to the superposition principle
\begin{align}
    &E_z(x_s,y_s,z=ct-s,t)= \\ \nonumber &E_{1,z}(x_s,y_s,z=ct-s,t)+E_{2,z}(x_s,y_s,z=ct-s,t)
\end{align}
is the sum of the losses of the second particle to the radiation 
\begin{align}
    U_{\mathrm{rad}}^{(2)}=c Q_1\int\limits_{-\infty}^{\infty}E_{2,z}(x_0,y_0,z=ct,t)dt.
\end{align}
and a wake field interaction term
\begin{align}
cQ_2\int\limits_{-\infty}^{\infty} &E_{1,z}(x_s,y_s,z=ct-s,t)dt=\\- &c Q_1Q_2 w_\parallel(s). \nonumber    
\end{align}
Here $w_\parallel(s)$ is the longitudinal wake potential.

As before, assuming $U_{\mathrm{rad}}^{(1,2)}<0$ and considering the inequality Eq.\eqref{eq:radfin2} 
%(which is also valid for the total flux), and assuming that the incoming pipe is enclosed by the outgoing pipe $D_A \subseteq D_B$, 
the following inequality for the wake potential is obtained
 \begin{align}
 \label{eq:radfinW}
     \phi_{B}^{\mathrm{cr}}-\phi_{A}^{\mathrm{cr}}<w(s)<2\phi_{B}^{\mathrm{cr}}-2\phi_{C}^{\mathrm{cr}},
 \end{align}
with $\phi^{cr}$ being the normalized ``cross'' flux though the corresponding cross section
\begin{align}
\phi^{\mathrm{cr}}_{A,B,C}=\frac{1}{4\pi }\int\limits_{-\infty}^{\infty}\int_{D_A,D_B,D_C}\frac{e_1e_2^*+e_2e_1^*}{Q_1Q_2}dxdydt.   
\end{align}
We note that the expression for the ``cross''-flux is in full agreement with the one from Ref.\cite{Opt1}

With the help of Eq.\eqref{eq:dpe} we evaluate the integral over $t$ and obtain for the cross section $A$ 

\begin{align}
\label{eq:redlf}
    &\phi^{\mathrm{cr}}_{A}= \frac{\delta(s)}{\pi } I_A.
\end{align}

Here $I_A$ is an exchange integral that reads:
\begin{align}\label{eq: exchange integral}
I_A=2\int\limits_{D_A}\Re\left\{\left[\frac{f'(\o,\o_0)}{f(\o,\o_0)} \right]^*\left[\frac{f'(\o,\o_s)}{f(\o,\o_s)} \right] \right\}dxdy.
\end{align}
The integral can be evaluated explicitly (see Appendix \ref{app:1} for the details) and reads
\begin{align}
    I_A=-\pi\left\{\ln\left[f(\o_0,\o_s)\right]+\ln\left[f(\o_s,\o_0)\right]\right\}.
\end{align}
Expression above for the integral can be simplified further to
\begin{align}
\label{eq:intres}
    I_A=-2\pi\ln \left|f(\o_s,\o_0)\right|.
\end{align}
Combining Eq.\eqref{eq:radfinW} with Eq.\eqref{eq:redlf} and accounting for the Eq.\eqref{eq:intres} we get
\begin{align}
\label{eq:wake gen}
    w_\parallel (s)&>2\delta(s) \ln \pskob{\frac{f(\o_s,\o_0)}{g(\o_s,\o_0)}}, \nonumber \\
    w_\parallel (s)&< 4\delta(s)\ln \left|\frac{q(\o_s,\o_0)}{g(\o_s,\o_0)}\right|.
\end{align}    
Here, as before, the function $g$ - is the conformal mapping of the cross section $D_B$ onto a unit disk, and function $q$ - is the conformal mapping of the cross section $D_C$ onto a unit disk. In all cases, the index $0$ corresponds to the leading particle and the index $s$ to the trailing particle.

Transverse wake potential is recovered from Eq.\eqref{eq:wake gen} with the help of the Panofsky-Wenzel theorem \cite{PW}
\begin{align}
\label{eq:PWg}
    \partial_s \vec{w}_\perp=\nabla_\perp w_\parallel.
\end{align}
In the complex notations equations above reads
\begin{align}
\label{eq:PWcm}
    \partial_s w^*_\perp=2\partial_{\omega_s} w_\parallel,
\end{align}
where $w_\perp=w_x+i w_y$. Combining Eq.\eqref{eq:wake gen} and \eqref{eq:PWcm} we arrive at the estimates for the transverse wake potential in a form 
\begin{align}
\label{eq:wake tr gen est}
    |w_x (s)|&>2\theta(s)\left| \Re \left\{ \left[\frac{f'(\o_s,\o_0)}{f(\o_s,\o_0)}-\frac{g'(\o_s,\o_0)}{g(\o_s,\o_0)}\right] \right\} \right|, \nonumber \\
    |w_y (s)|&>2\theta(s)\left| \Im \left\{ \left[\frac{f'(\o_s,\o_0)}{f(\o_s,\o_0)}-\frac{g'(\o_s,\o_0)}{g(\o_s,\o_0)}\right] \right\} \right|, \nonumber \\
    |w_x (s)|&< 4\theta(s)\left| \Re \left\{\left[\frac{q'(\o_s,\o_0)}{q(\o_s,\o_0)}-\frac{g'(\o_s,\o_0)}{g(\o_s,\o_0)}\right]\right\} \right|, \\
    |w_y (s)|&< 4\theta(s)\left| \Im \left\{\left[\frac{q'(\o_s,\o_0)}{q(\o_s,\o_0)}-\frac{g'(\o_s,\o_0)}{g(\o_s,\o_0)}\right]\right\} \right|. \nonumber
\end{align} 
Here $\theta(s)$ is the Heaviside step-function. 

Note that the lower and the upper bounds always have the same sign, so the sign of the wake is fixed by the sign of the corresponding upper bound, which means that the function $\frac{q'(\o_s,\o_0)}{q(\o_s,\o_0)}-\frac{g'(\o_s,\o_0)}{g(\o_s,\o_0)}$ gives the structure of the transverse wake potential. This fact can be used to construct a two-dimensional kick map estimate based solely on the complex derivative of the upper bound in Eq.\eqref{eq:wake gen}. Thus, we introduce a bounding transverse wake potential of the form 
\begin{align}
\label{eq:wake tr gen}
    \mathcal{W}_\perp (s)^*=4\theta(s)\left[\frac{q'(\o_s,\o_0)}{q(\o_s,\o_0)}-\frac{g'(\o_s,\o_0)}{g(\o_s,\o_0)}\right].
\end{align}
The above expression makes it possible to estimate the maximum amplitude and the structure of the transverse wake potential at the same time.

%==============================================================
%==============================================================

\section{Examples}

%==============================================================
%==============================================================

In this section, we consider two basic conformal maps and apply them to evaluate the wake potential for round-to-round, flat-to-flat, and flat-to-round step-out transitions.

\subsection{Conformal maps}

\begin{figure}[t]
    \includegraphics[width=0.45\textwidth]{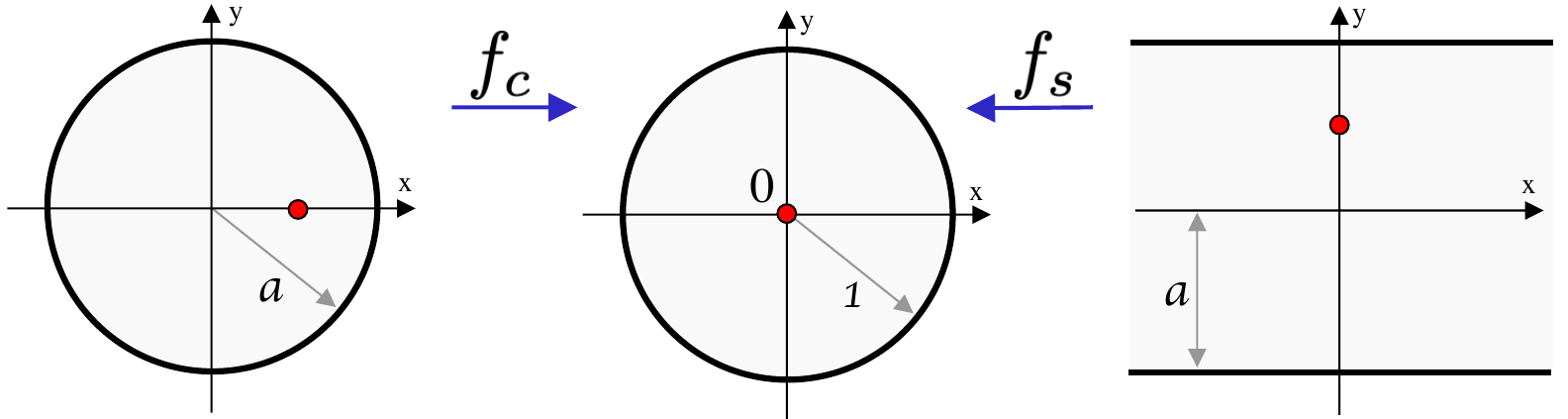}
    \caption{Diagrams of conformal mappings. Displaced disk on a unit disk (left part of the diagram) and strip parallel to the $x$-axis on a unit disk (right part of the diagram).}
    \label{Fig:conf}
\end{figure}

The conformal map $f_{c(a)}(\o,\o_i)$ of a disk (see Fig.\ref{Fig:conf} left part of the diagram) with radius $a$ onto a unit disk such that the point of a source $\o_0$ is mapped to the center of a unit disk is given by \cite{shabat,silverman}
\begin{equation}\label{eq: chi for arb circle}
    f_{c}(\o,\o_0)=a\frac{\o-\o_0}{a^2-\o\o_0^*}
\end{equation} 

A function $F(\o)$, which maps an infinite parallel to the $y$-axis strip with the height $2a$ to a disk of radius $a$ is given by \cite{shabat,silverman}: 
\begin{align}\label{eq:strp}
    F(\o)=a\tan\left[\frac{\p}{4a}\o\right].
\end{align}
Combining \eqref{eq: chi for arb circle} and \eqref{eq:strp} we arrive at the map for the strip onto a unit disk with the point $\omega_0$ mapped to the center of the unit disk 
\begin{align}
f(\omega,\omega_0)=\frac{\tan\left(\frac{\pi}{4}\frac{\omega}{a} \right)-\tan\left(\frac{\pi}{4}\frac{\omega_0}{a} \right)}{1-\tan\left(\frac{\pi}{4}\frac{\omega}{a} \right)\tan\left(\frac{\pi}{4}\frac{\omega_0^*}{a} \right)}.
\end{align}
Substitution $\omega \to i\omega^*$ makes the strip parallel to the $x$-axis (see Ref.\cite{planar}) and we finally get (see Fig.\ref{Fig:conf} right part of the diagram)
\begin{align}
\label{eq:conf_s}
f_{s(a)}(\omega,\omega_0)&=i\frac{\tanh\left(\frac{\pi}{4}\frac{\omega^*}{a} \right)-\tanh\left(\frac{\pi}{4}\frac{\omega_0^*}{a} \right)}{1-\tanh\left(\frac{\pi}{4}\frac{\omega^*}{a} \right)\tanh\left(\frac{\pi}{4}\frac{\omega_0}{a} \right)}.
\end{align}

\begin{figure*}
    \includegraphics[width=\textwidth]{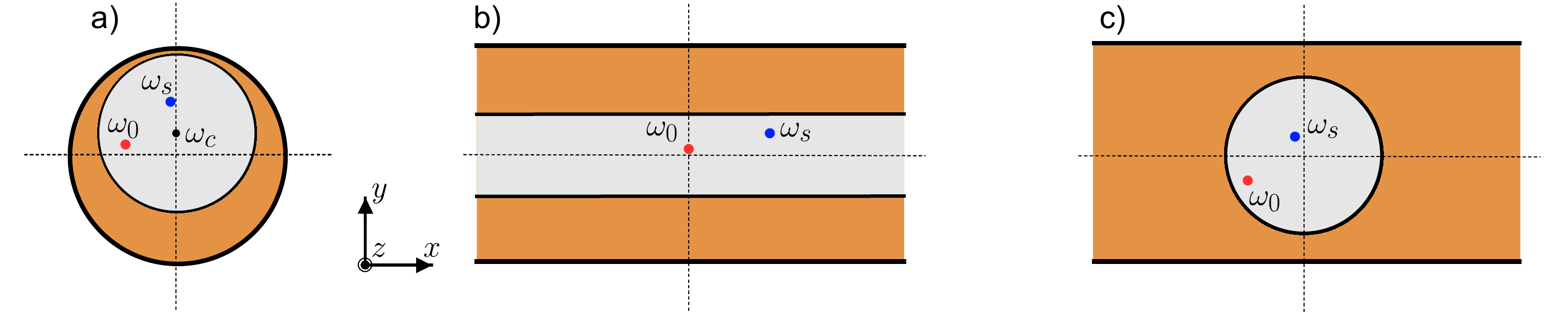}
    \caption{Sketches of the several transitions that was examined. Orange area corresponds to the cross section of the ``out" waveguide, and light grey to the cross section of the ``in" waveguide. The red dots are for the source particle, the blue dots are for the test particle, and the point $\o_c$ corresponds to the center of the incoming pipe.}
    \label{fig: examples of transitions}
\end{figure*}

%==============================================================
\subsection{Step-out transitions}
%==============================================================

We assume that the cross sections of the incoming waveguide and the collimating section are equal $D_C\equiv D_B$ ($q\equiv g$).
%, and the cross section of the outgoing waveguide encloses the cross section of the incoming pipe $D_A \subset D_B$. 
Such a transition is called step-out.

For the step-out transition as well as for the collimator with such a step-out transition, the upper bound of the longitudinal wake potential is given by Eq.\eqref{eq:wake gen} and the transverse wake potential can be estimated using Eq.\eqref{eq:wake tr gen}.

%==============================================================
 \subsubsection{Round to round transition}
%==============================================================

We examine the transition between two round pipes with radii $a_1$ and $a_2$ respectively (Fig.\ref{fig: examples of transitions}a), the symmetry axis of the first (incoming) pipe is at the point $\o=\o_c$, and the symmetry axis of the second (outgoing) pipe is at $\o=0$. Since we are studying a step-out transition, we introduce a constraint on the pipe parameters: $|\o_c|+a_1<a_2$. The cross section of the transition is sketched in Fig.\ref{fig: examples of transitions}(a).

With the help of Eq.\eqref{eq: chi for arb circle} we calculate the conformal maps $q$ and $g$:
\begin{align}\label{eq: chis for pipe to pipe}
&q=f_{c(a_1)}(\o-\o_c,\o_{0,s}-\o_c)= \\ \nonumber &\frac{a_1(\o-\o_{0,s})}{a_1^2-(\o-\o_c)(\o_{0,s}^*-\o_c^*)},\\
&g=f_{c(a_2)}(\o,\o_{0,s})=\frac{a_2(\o-\o_{0,s})}{a_2^2-\o\o_{0,s}^*}. 
\end{align}

We combine Eq.\eqref{eq: chis for pipe to pipe} and Eq.\eqref{eq:wake gen}, and arrive at the upper bound for the wake potential in the form:
\begin{equation}\label{eq: wake for arbitrary pipe to pipe}
w_\parl<4\d(s)\ln\pskob{\frac{a_1}{a_2}\frac{a_2^2-\o_0^*\o_s}{a_1^2-(\o_s-\o_c)(\o^*_0-\o^*_c)}}.
\end{equation}

In the trivial case, when both waveguides and particles are aligned to the symmetry axis ($\o_c=\o_0=\o_s=0$) Eq.\eqref{eq: wake for arbitrary pipe to pipe} simplifies to a well known expression \cite{HfKf,Opt2}
\begin{align}
\label{eq:cyl_los}
    w_\parl<4\d(s)\ln\frac{a_2}{a_1}.
\end{align}

If we consider a slightly misaligned waveguide ($\o_c \ll a_1$) and a small displacement of both particles from the origin ($\pskob{\o_{0,s}}\ll a_{1}$), then the 
Taylor expansion to the lowest order in $|\omega_{0,c}|$ of the Eq.\eqref{eq: wake for arbitrary pipe to pipe} reads:
\begin{align}
  w_\parl=4\d(s)\ln\left |\frac{a_2}{a_1}-\frac{a_2|\o_0||\o_c|}{a_1^3}\cos\left(\phi_0-\phi_c \right) \right|.  
\end{align}

For the transverse wake potential with the help of the Eq.\eqref{eq:wake tr gen} and Eqs.\eqref{eq: chis for pipe to pipe} we get 
\begin{align}
    \mathcal{W}_\perp^* (s)&= \\ \nonumber &4\theta(s)\left\{\frac{\o_0^*\left[a_2^2-a_1^2-\o_c\o_0^*\right]-\o_c^*\left[a_2^2-\o_c\o_0^* \right]}{\left[a_2^2-\o_s\o_0^* \right] \left[a_1^2-(\o_s-\o_c)(\o_0^*-\o_c^*) \right]}\right\}.
\end{align}
Assuming $\o_c=0$ and $\o_s=\o_0\ll a_{1,2}$ we recover a known formula derived in Ref.\cite{dp}.
\begin{align}
\label{eq:tr_cyl}
    \mathcal{W}_\perp (s)=4\theta(s)\o_0\left[\frac{1}{a_1^2}-\frac{1}{a_2^2} \right].
\end{align}

%==============================================================
\subsubsection{Planar to planar transition}
%==============================================================

Next we consider a transition from a planar aperture with half the gap $a_1$ to a planar aperture with half the gap $a_2$. As before $a_2>a_1$. The cross section of the transition is shown in Fig.\ref{fig: examples of transitions}b.

Simplification of the Eq.\eqref{eq:conf_s} gives
\begin{equation}\label{eq: chis for planar to planar}
\begin{aligned}
&q=f_{s(a_1)}=i\frac{\sinh \left[\frac{\pi}{4}\frac{x-i(y-y_0)}{a_1}\ \right]}{\cosh \left[\frac{\pi}{4}\frac{x-i(y+y0)}{a_1} \right]},\\
&g=f_{s(a_2)}=i\frac{\sinh \left[\frac{\pi}{4}\frac{x-i(y-y_0)}{a_2}\ \right]}{\cosh \left[\frac{\pi}{4}\frac{x-i(y+y0)}{a_2} \right] }.
\end{aligned}
\end{equation}
Above, we assumed that without the loss of generality $\omega_0=iy_0$ and $\omega_s=x+iy$.
We combine Eq.\eqref{eq:wake gen} and Eq.\eqref{eq: chis for planar to planar} and arrive at the expression for the upper bound on the $w_\parl$ in the form
\begin{align}
    &w_\parl<4\d(s)\times \\ \nonumber&\ln\pskob{\frac{\sinh \left[\frac{\pi}{4}\frac{x-i(y-y_0)}{a_1}\ \right]}{\cosh \left[\frac{\pi}{4}\frac{x-i(y+y0)}{a_1} \right]} \frac{\cosh \left[\frac{\pi}{4}\frac{x-i(y+y_0)}{a_2}\ \right]}{\sinh \left[\frac{\pi}{4}\frac{x-i(y-y0)}{a_2} \right]}}.
\end{align}

If both particles are at the origin $\omega_s=\omega_0=0$ we recover a known result \cite{Opt2} that coincides with Eq.\eqref{eq:cyl_los} 
\begin{align}
w_\parl<4\d(s)\ln\frac{a_2}{a_1}.
\end{align}
The transverse wake potential is calculated based on the Eq.\eqref{eq:wake tr gen} and Eqs.\eqref{eq: chis for planar to planar}. After some algebra we arrive at
\begin{align}
    \mathcal{W}_\perp^* (s)&= \frac{4\theta(s) i\pi \cos\left[ \frac{\pi y_0}{2 a_1} \right]}{2a_1 \sin \left[\frac{\pi y_0}{2 a_1} \right]+2ia_1\sinh\left[\frac{\pi}{2a_1}\left( x+iy \right) \right]}- \nonumber \\ &\frac{4\theta(s)i\pi \cos\left[ \frac{\pi y_0}{2 a_2} \right]}{2a_2 \sin \left[\frac{\pi y_0}{2 a_2} \right]+2ia_2\sinh\left[\frac{\pi}{2a_2}\left( x+iy \right) \right]} .
\end{align}
Assuming $x=0$ and $y\ll a_{1,2}$ and $y_0 \ll a_{1,2}$ we get
\begin{align}
    \mathcal{W}_y (s)=4\theta(s)\frac{\pi^2}{4}\left[\frac{y}{6}+\frac{y_0}{3} \right]\left[\frac{1}{a_1^2}-\frac{1}{a_2^2} \right].
\end{align}
For a special case of $y=y_0$, we arrive at the following expression
\begin{align}
    \mathcal{W}_y (s)=4\theta(s)\frac{\pi^2}{8}y_0\left[\frac{1}{a_1^2}-\frac{1}{a_2^2} \right],
\end{align}
that differs from the Eq.\eqref{eq:tr_cyl} by a factor of $\pi^2/8$.

%==============================================================
\subsubsection{Round to planar transition}
%==============================================================

Finally, consider the transition from the pipe with radius $a_1$ centered at the origin to the planar waveguide of an aperture $2a_2$ (Fig.\ref{fig: examples of transitions}c). We assume $a_1<a_2$.

With the help of the Eq.\eqref{eq: chi for arb circle} and Eq.\eqref{eq:conf_s}, we calculate conformal maps $q$ and $g$:
\begin{align}\label{eq: chis for pipe to planar}
&q=f_{c(a_1)}(\o,\o_{0,s})=\frac{a_1(\o-\o_{0,s})}{a_1^2-\o \o_{0,s}^*},\\
&g=f_{s(a_2)}(\o,\o_{0,s})
= \\ \nonumber &i\frac{\tanh\left(\frac{\pi}{4}\frac{\omega^*}{a_2} \right)-\tanh\left(\frac{\pi}{4}\frac{\omega^*_{0,s}}{a_2} \right)}{1-\tanh\left(\frac{\pi}{4}\frac{\omega^*}{a_2} \right)\tanh\left(\frac{\pi}{4}\frac{\omega_{0,s}}{a_2} \right)}.
\end{align}
Substituting Eq.\eqref{eq: chis for pipe to planar} into Eq.\eqref{eq:wake gen} and proceeding to the limit $\omega_s=\omega_0=0$ we get
\begin{equation}
w_\parl<4\d(s)\ln\left[\frac{4}{\pi} \frac{a_2}{a_1} \right].
\end{equation}
We note that along with the familiar term $\ln \frac{a_2}{a_1}$ as in eq.\eqref{eq:cyl_los} we get an additional $\ln \frac{4}{\pi}$ - a form factor mismatch contribution. Note that geometry mismatch tends to increase the losses. 

Substituting $\omega_0=x_0+iy_0$ and $\omega_s=x+iy$ into Eq.\eqref{eq: chis for pipe to planar} with the help of Eq.\eqref{eq:PWg} and Eq.\eqref{eq:wake gen} we get
\begin{align}
    &\mathcal{W}_x (s)=4\theta(s)\frac{x_0}{a_1^2}, \\ \nonumber
    &\mathcal{W}_y (s)=4\theta(s)\frac{y_0}{a_1^2}\left[1-\frac{\pi^2}{8}\frac{a_1^2}{a_2^2} \right].
\end{align}
Here we have assumed $y=y_0$ and $x=x_0$ in the final expression for a lowest-order Taylor decomposition.
Note that the transverse wake has the same structure as in the round-to-round and planar-to-planar cases and consists of a difference of the inverse squares of the corresponding characteristic sizes in the corresponding direction. Each characteristic size is multiplied by a corresponding form factor. For the $\mathcal{W}_y$ it is the same as for the planar case - $\pi^2/8$. For $\mathcal{W}_x$, the $x$ parameter of the planar structure is $\infty$, and only the first term survives. 

Interestingly, if $a_2^2=\pi^2/8a_1^2$ the $y$ component of the transverse wake potential vanishes $\mathcal{W}_y=0$. Furthermore, if $a_1^2<a_2^2<\pi^2/8a_1^2$ we get $\mathcal{W}_y<0$ and the wake is focusing. The maximum focusing strength is at $a_1=a_2$ 
\begin{align}
    \mathcal{W}_y (s)<-4\theta(s)\frac{y_0}{a_1^2}\left[\frac{\pi^2}{8}-1\right]\approx-0.93\theta(s)\frac{y_0}{a_1^2}.
\end{align}
We can see that the strength of the focusing force is about $1/4$ of the kick in the $x$ direction.

%==============================================================
%==============================================================
\section{Conclusion}
%==============================================================
%==============================================================

We have revised the evaluation of the geometric wake potential in the collimators and step transitions with perfectly conducting walls. Building on the model and ideas of Ref.\cite{HfKf} and Ref.\cite{mySTAB}, we demonstrated that the wake potential can be well estimated using the conformal mapping technique. In the case where the cross sections of the incoming collimating and outgoing pipes have simple shapes (rectangular, slab, circular, elliptical), the mapping is evaluated explicitly. This in turn allows convenient analytical calculation of the corresponding form factors for both longitudinal and transverse wake potentials. If the cross section is an arbitrary single connected region, the corresponding map and its derivative can be numerically evaluated using the Zipper algorithm \cite{Zipper}. Zipper allows fast and accurate evaluation of the conformal mapping, but more importantly, it can produce the derivative of the mapping along with the map without loss of accuracy and increase in computational time.

We observe that if $g$ and $f$ are the same mappings (i.e., differ only by the scaling parameter) of a symmetric cross section (a pipe with the axis of zero transverse wake), the step-out wake for the beam on that axis is always $\ln b/a$. However, the transverse wake will always differ by a form factor from the corresponding expression for the transition between round pipes. 

The bounding expressions for the longitudinal Green's function Eq.\eqref{eq:wake gen} and for the transverse Green's function Eq.\eqref{eq:wake tr gen est} can be applied to the exploration of the appropriate transverse beam shape to minimize or maximize the corresponding wake effect.

%We point out that in more complex transitions, where one pipe is not enclosed by another, one can extend the presented ideas once the reflected flux is properly estimated. It is believed that the inequalities given by Eq.\eqref{eq:wake gen} and Eq.\eqref{eq:wake tr gen} are general once the corresponding effective cross sections are defined based on the estimation of the reflected flux.

%\newpage

%==============================================================
%==============================================================

\appendix

%==============================================================
\section{Evaluation of the cross flux integral}
\label{app:1}
%==============================================================

\subsection{Approach 1}

We start with the new notation for the conformal mapping function
\begin{equation}\label{eq: new notation}
    f\skob{\o,\o_i}\equiv\chi_i(\o).
\end{equation}
In this section $i,j=0,s$ and $\o_i$ is the coordinate of the corresponding particle. 

We consider the Stock's theorem in a complex plane, that reads
\begin{equation}\label{eq: from dx to dw}
    \iint_Dgdr^2=\frac{1}{2i}\oint_{\partial D}\int g d\o^*d\o.
\end{equation}
It holds for an arbitrary complex function $g$.

Before we may use this formula, we must properly define the integration domain. We cut infinitesimally small circles with a radius $r_0$ centered on $\o_0$ and $\o_s$. We denote these cut-out areas $D_0$ and $D_s$ respectively and define the integration domain as : $D=D_{A,B}\setminus(D_{0}\cup D_s)$

With Eq.\eqref{eq: new notation} under Eq.\eqref{eq: from dx to dw} expression for the exchange integral Eq.\eqref{eq: exchange integral} gets the following form
\begin{equation}\label{eq: 2 particles energy}
\begin{gathered}
        \frac{1}{2i}\oint_{\partial{D}} \int\skob{\frac{\chi_0'}{\chi_0}\frac{\chi_s'^*}{\chi_s^*}+\frac{\chi_s'}{\chi_s}\frac{\chi_0'^*}{\chi_0^*}}d\o^*d\o =\\=\frac{1}{2i}\oint_{\partial{\til{D}'}}\skob{\ln\chi_s^* d\ln\chi_0+\ln\chi_0^* d\ln\chi_s}
\end{gathered}\end{equation}

The integral over $\partial \til{D}'$ is a linear combination of integrals over $\partial  D_0$, $\partial  D_s$ and $\partial D$ (see Fig.\ref{fig: contour}a).

First, we evaluate the integral on the contours $\partial D_i$ ($i=0,s)$. On these contours we expand $\chi_i\approx \chi_i'(\o_i)r_oe^{i\ph}$, and get: 
\begin{align}
    &\ln \skob{\chi_i}= \ln\pskob{r_0\chi_i'(\o_i)}+i\ph+O(r_0) \\ \nonumber
    &\ln \skob{\chi_j}= \ln \skob{\chi_j(\o_i)}+\frac{\chi_j'(\o_i)}{\chi_j(\o_i)}r_oe^{i\ph}+O(r_0)
\end{align}
Substitution of this expansion into \eqref{eq: 2 particles energy} gives:
\begin{align}
    &\frac{1}{2i}\int_{2\p }^{0}\left[ \ln \skob{\chi_j(\o_i)}+\frac{\chi_j'(\o_i)}{\chi_j(\o_i)}r_oe^{i\ph} \right]id\ph+\nonumber\\
    &\frac{1}{2i}\int_{2\p }^{0}\left[ \ln\pskob{r_0\chi_i'(\o_i)}+i\ph \right]\frac{\chi_j'(\o_i)}{\chi_j(\o_i)}r_oe^{i\ph}id\ph= \\ \nonumber &-\p \ln \skob{\chi_j(\o_i)}+O(r_0)
\end{align}

The integral of $\partial D_i$ is computed in a similar fashion, with the interchange $i\leftrightarrow j$. 
The remaining integral over $\partial D$ is equal to zero due to the boundary condition Eq.\eqref{eq:Php}. 

Combining all together we get the final expression for the exchange integral:
\begin{equation}
    I=-\pi\left\{\ln\left[f(\o_0,\o_s)\right]+\ln\left[f(\o_s,\o_0)\right]\right\}.
\end{equation}

%==============================================================
\subsection{Approach 2}
%==============================================================

To calculate this integral in more rigorous way we introduce new conformal map $\Tilde{\chi}_s$ that is based on $\chi_0$
\begin{equation}\label{eq:app: chi til s}
\til{\chi}_s(\o)=\frac{\chi_0(\o_s)-\chi_0(\o)}{1-\chi_0(\o_s^*)\chi_0(\o)}.
\end{equation}

We treat $\chi_s$ as a function of $\chi_0$ and move from integration over $d\o d\o^*$ to integration over $d\chi_0 d\chi_0^*$. After we substitute Eq.\eqref{eq:app: chi til s} and perform a change of variables ($\pskob{\partial\c_0/\partial\o}^2d\o d\o^*=d\c_0d\c_0^*$) the integral \eqref{eq: exchange integral} reads
\begin{align}
     \frac{1}{2i} \oint_{\partial\til{D}'} d\chi_0 \int d\chi_0^* \left[\frac{1}{\chi_0}\skob{\frac{\Tilde{\chi}_s'}{\Tilde{\chi}_s}}^*+\frac{1}{\chi_0^*}\frac{\Tilde{\chi}_s'}{\Tilde{\chi}_s} \right].
\end{align}
Here, the derivative $\Tilde{\chi}_s'$ is taken with respect to the variable $\chi_0$. We substitute $\Tilde{\chi}_s$ with the explicit expression and get 
\begin{equation}\label{eq: app simplified int}
\begin{gathered}
   \frac{1}{2i}\oint_{\partial\til{D}'} d\chi_0 \left \{ \frac{1}{\chi_0}\left[\ln\skob{\chi_0^*-\chi_0(\o_s)}-\ln\skob{1-\chi_0^*\chi_0(\o_s)}\right]+\right.\\ \left. +\ln\skob{\chi_0^*}\frac{1-\chi_0(\o_s)^2}{(\chi_0(\o_s)-\chi_0)(1-\chi_0(\o_s)\chi_0)} \right \}.
\end{gathered}
\end{equation}

\begin{figure}[t]
\includegraphics[width=0.3\textwidth]{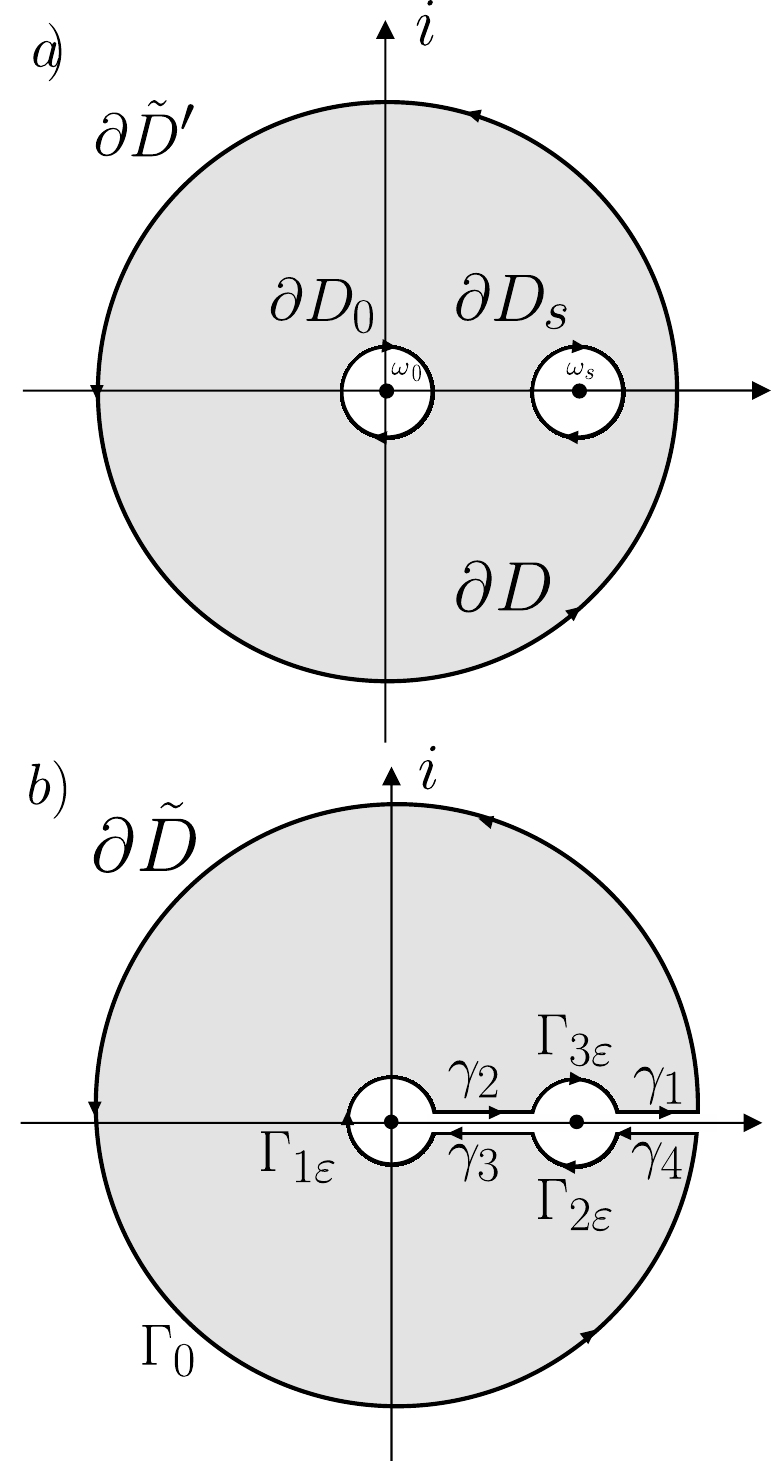}    \caption{Sketch of the integration contour}\label{fig: contour}
\end{figure}

To this end present approach does not differ from the Approach 1. The rigor comes into play now. 
It is known that the function $\ln\skob{z}$ is not analytic and has a cut in the complex plane. This complicates the evaluation of the integral. To perform this evaluation with care, we introduce the new contour $\partial \tilde{D}$ that is sketched in Fig.\ref{fig: contour}b.

Integral over $\partial \til{D}$ is zero, as the function under integral does not have any singularities or cuts in the region enclosed by the contour. Integration over $\partial\til{D}'$ corresponds to the integration over $\G_{1\e}+\G_{2\e}+\G_{3\e}+\G_{0}$, but this integral equals minus the integral over $\g=\g_1+\g_2+\g_3+\g_4$. The integral of the contour $\g_1+\g_2+\g_3+\g_4$ is evaluated as follows.  We introduce the parametrization: on contour $\g_1+\g_2$ --  $\chi_0=x$; on contour $\g_1+\g_2$ -- $\chi_0=xe^{2\pi i}$. With $x$ in the range: $\skob{0;\chi_0(\o_s)-\e}\cup\skob{\chi_0(\o_s)+\e;1}$, but in different directions.

Incorporating everything in Eq.\eqref{eq: app simplified int} we get:
\begin{align}
    I&=\frac{1-\chi_0(\o_s)^2}{2i}v.p.\int_0^1  \frac{2\pi i}{\skob{\chi_0(\o_s)-x}\skob{\chi_0(\o_s)x-1}}dx \nonumber\\
    &=-2\pi\ln{\chi_0(\o_s)}
\end{align}

This expression is equivalent to
\begin{align}
        I&=-\pi\left[\ln\chi_0(\o_s)+\ln\chi_s(\o_0)\right]
\end{align}
once we recall that $\chi_0(\o_s)=\til{\chi}_s(\o_0)$. The latter follows from the straightforward substitutions in Eq.\eqref{eq:app: chi til s}.

\section{Connection between magnetic and electric fields}
\label{app:2}

Here we provide a detailed explanation on connection between complex functions that correspond to electric and magnetic fields in Eq.\eqref{c_fl} in the cross section of the waveguide, namely $h=ie$. 

One can show that  Eq.\eqref{Max3} in terms of complex functions Eq.\eqref{c_fl} reads
\begin{equation}\label{eq: app 2 nabla fields}
\begin{aligned}
\nabla_c^* e & =\left(4\pi\rho+\frac{\partial E_z}{\partial \zeta }-i\frac{\partial H_z}{\partial \zeta }\right), \\
\nabla_c^* h & =i\left(4\pi\rho+\frac{\partial E_z}{\partial \zeta }-i\frac{\partial H_z}{\partial \zeta }\right). \nonumber
\end{aligned}
\end{equation}
Under the assumption of no radiation we get
\begin{equation}\label{eq: app 3 nabla fields}
\begin{aligned}
\nabla_c^* e & =4\pi\rho, \\
\nabla_c^* h & =i4\pi\rho. \nonumber
\end{aligned}
\end{equation}
We introduce operator
\begin{equation}
\nabla_c^*=\frac{\partial}{\partial x}-i \frac{\partial}{\partial y}=2\frac{\partial}{\partial\omega}
\end{equation}

From Eq.\eqref{eq: app 3 nabla fields} it is clear that $h$ and $ie$ may differ only by some analytical function $g(w)$. 

In order to define this function, we have to examine boundary conditions. We examine PEC waveguide that corresponds to following boundary conditions
\begin{equation}
\begin{aligned}
    \vc{n}\times\vc{E}=0 \rightarrow \Re\sqskob{e\tau*}=0,\\
        (\vc{n},\vc{H})=0 \rightarrow \Im\sqskob{h\tau*}=0\nonumber.
\end{aligned}
\end{equation}
From these conditions we conclude that $g(\omega)$ must to be 0 at the boundary of the waveguide. We recall that $g(\omega)$ is an analytical function and reaches its maximum and minimum at the boundary. From the latter we conclude that $h=ie$.

\begin{acknowledgments}
The author thanks Prof. Donald Marshall of George Washington University for his help with the Zipper algorithm. The work was supported by the Foundation for the Advancement of Theoretical Physics and Mathematics "BASIS" $\#$22-1-2-47-17 and the ITMO Fellowship and Professorship Program.
\end{acknowledgments}

\bibliographystyle{ieeetr}
\bibliography{refs}

\end{document}